\documentclass{article}
\hoffset= - 1.0in         
\usepackage{amssymb}
\usepackage{amsmath}

\def\be{\begin{equation}}
\def\ee{\end{equation}}

\title{\Large {\bf Geodesic deviation and particle creation in curved spacetimes}
\vspace{.2cm}}
\author{{\bf A. Mironov}\footnote{ {\small {\it
Lebedev Physics Institute} and {\it ITEP, Moscow, Russia}};
mironov@itep.ru; mironov@lpi.ru}, {\bf A. Morozov}\thanks{{\small
{\it ITEP, Moscow, Russia}}; morozov@itep.ru} \ and {\bf
T.N. Tomaras}\thanks{{\small {\it Department of Physics and ITCP,
University of Crete};
tomaras@physics.uoc.gr}}\date{ }}

\begin{document}

\maketitle

\vspace{-6.0cm}

\begin{center}
\hfill FIAN/TD-15/11\\
\hfill ITEP/TH-25/11\\
\hfill CCTP-2011-26  \\
\end{center}

\vspace{4.5cm}

\begin{abstract}
A quantum mechanical picture, relating accelerated geodesic deviation to creation
of massive particles via quantum tunneling in curved background spacetimes, is presented. The effect is 
analogous to pair production
by an electric field and leads naturally to production of massive particles in de Sitter
and superluminal FRW spacetimes. The probability of particle production in de Sitter space per unit volume 
and time is computed in a leading semiclassical approximation and shown to coincide
with the previously obtained expression. 
\end{abstract}

\bigskip

\bigskip

{\bf 1.} When embedded into an external background, the ``vacuum" of
quantum field theory can lose stability because of particle creation. The difference from classical or quantum
radiation is that the particles are produced through quantum
tunneling.

The standard paradigm is the ``Klein paradox" \cite{KP}, i.e. creation of electron-positron pairs by an
electric field $E$. Let us for simplicity consider a constant
electric field along the $z-$axis in a capacitor of width $L$. 
The probability to create a single particle-antiparticle pair (the Schwinger process \cite{Schwinger}) in its center 
of mass frame
is given by the tunneling exponential $\left| \exp(i\int p(z)\,dz)\right|^2$, where the momentum $p(z)$
is the imaginary solution of the dispersion relation
\footnote{
The energies ${\cal E}_1, {\cal E}_2$ and momenta $p_1, p_2$ along the $z-$axis of the particles in the 
pair are related by
$$
\left[{1\over 2}\left({\cal E}_+-eEz_-\right)\pm{1\over 2}\left({\cal E}_--eEz_+\right)\right]^2=
\left[{1\over 2}\left(p_+\pm p_-\right)\right]^2+m^2\,,
$$
with $z_\pm=z_1\pm z_2$, ${\cal E}_\pm={\cal E}_1\pm {\cal E}_2$, $p_\pm=p_1\pm p_2$. Adding these
two equations one obtains
$$
\left({\cal E}_+-eEz_-\right)^2+\left({\cal E}_--eEz_+\right)^2=p_+^2+p_-^2+4m^2\,,
$$
which in the center of mass frame $p_+=0$, $z_+=0$ and ${\cal{E}}_-=0$, reduces to (\ref{dr}).
}
\be
\label{dr}
\Big({\cal E}_+-eA_0(z)\Big)^2 = 4p^2 + 4m^2\,,
\ee
with the total energy of the pair ${\cal E}_+=-2m$. In the above equation $z\equiv z_1-z_2$ is the 
distance of two particles, $p\equiv (p_1-p_2)/2$ is its conjugate momentum, while $A_0(z) = \int^z Edz$.

The characteristic parameter is the ``creation length" $l$,
defined by the condition that the energy offered by the electric field compensates for the rest energy
on the particles produced, i.e. $eA_0(l) = e \int_0^l Edz = 2m$, which gives $l=2m/eE$ for a constant field $E$.
The pair creation occurs only if $L>l$ and the probability of creation per unit volume and
unit time interval is the well known exponential \cite{Schwinger}
\be
\label{proel}
\theta(L-l) \exp\left(-\int_0^l \sqrt{4m^2 - e^2 E^2z^2}\ dz\right) =
\theta (L-l) e^{-\pi m^2/eE} \,.
\ee
This integral runs over the distance $z$ between the particles in the pair and
describes the tunneling from the point $z=0$ at the ``moment" of creation to $z=l$. Note that the
applicability of the quasiclassical approximation requires the exponent in (\ref{proel})
to be large and negative, so that the exponential is small.

\vspace{0.5cm}

{\bf 2.} The situation becomes considerably more interesting when the electric field is replaced by a
gravitational one \cite{FT}. For instance, contrary to the case of a constant electric field, which can 
pull apart the two virtual particles of the pair, the equivalence principle forbids pair creation in a {\it constant} 
gravitational field.

However, it is natural to expect that an analogous ``pulling-apart force" will occur also in gravity,
in the presence of curvature, as long as the latter has the right properties, which make close geodesics
diverge.

The equation of geodesic deviation in a given background specetime is \cite{LL2}
\be\label{gde}
D^2_\tau \eta^\mu=R^\mu_{\;\;\nu\rho\sigma}u^\nu u^\rho\eta^\sigma \,,
\ee
where $R^\mu_{\;\;\nu\rho\sigma}$ is the Riemann curvature tensor, $u^\mu=\partial x^\mu/\partial \tau$ 
is the four velocity along the
``central geodesic" of the one-parameter congruence $x^\mu(\tau, \lambda)$, parametrized by $\lambda$,
and $\eta^\mu=\partial x^\mu/\partial \lambda$, is essentially the coordinate distance between two close 
probe particles moving along the corresponding
geodesics. $D_\tau$ denotes the covariant differentiation with respect to the affine parameter $\tau$ along the 
central trajectory.

Consider the case of a maximally symmetric background spacetime with constant curvature scalar $R$.
De Sitter and AdS are two obvious examples of interest. The Riemann tensor of such a space is of the form
\be
\label{curv}
R_{\mu\nu\rho\sigma}=-{R\over 12}\left(g_{\mu\sigma}g_{\nu\rho}-g_{\mu\rho}g_{\nu\sigma}\right)
\ee
Inserting (\ref{curv}) into (\ref{gde}), one obtains
\be
\label{ddeta1}
D^2_\tau \eta^{\,\mu}=-{R\over 12}\left(\eta^{\,\mu}-(u\eta)u^{\,\mu}\right)\,.
\ee
In the study of the particle creation process, in particular, one can choose $\eta^\mu$ such that $(u\eta)=0$.
That is, in the rest frame of one of the particles with $u=(1,0,0,0)$, one observes the motion of the second one
at a proper space-like distance from the first, i.e. with $\eta^0=0$.
Then, equation (\ref{ddeta1}) simplifies to
\be
\label{7}
D^2_\tau \eta^{\,i}=-{R\over 12}\eta^{\,i} \,,
\ee
where Latin indices denote the spatial coordinates.

In eqs.(\ref{gde})-(\ref{7}) $\eta^\mu$ can not be chosen arbitrarily:
it is a tangent vector to a family of geodesics and
is related by the zero-curvature condition to $u^\mu$.
Physically significant is, however, not $\eta^\mu$,
but its length $z=||\eta||$, which measures the {\it distance}
between the geodesics.
In any spatially isotropic spacetime one may choose $\eta^i$ to have only one non-zero component.
In this case, using (\ref{7}) one obtains for the equation of motion of $z$
\footnote{Indeed,
$$D_\tau^2 z^2=2g_{\mu\nu}\eta^\mu D_\tau^2\eta^\nu+2g_{\mu\nu} D_\tau\eta^\mu D_\tau\eta^\nu=
-\frac{R}{6} g_{\mu\nu}\eta^\mu \eta^\nu+2g_{\mu\nu} D_\tau\eta^\mu D_\tau\eta^\nu=
-\frac{R}{6} z^2 +2g_{\mu\nu} D_\tau \eta^\mu D_\tau \eta^\nu\,.$$

On the other hand,
$$D_\tau^2 z^2=2z\frac{\partial^2 z}{\partial \tau^2}+2 \left(\frac{\partial z}{\partial\tau}\right)^2\,.$$

Therefore,
$$\frac{\partial^2 z}{\partial\tau^2}=-\frac{R}{12} z+ X ,$$ where
$$X=-\frac{1}{2z^3} (\eta_\mu D_\tau\eta^\mu \eta_\nu D_\tau\eta^\nu - z^2 g_{\mu\nu}D_\tau\eta^\mu D_\tau\eta^\nu)\,.$$
For any spacelike vector $\eta^\mu$ with only one non-zero component $X=0$.

A simple example to illustrate these points is $S^2$. For meridians $\phi=const$ on a
two-dimensional sphere with the metric $d\theta^2 + \sin^2\theta d\phi^2$
the role of affine parameter $\tau$ is played by $\theta$, and
the two tangent vectors are $u = (1,0)$ and $\eta = (0,1)$,
while $z = ||\eta|| = \sin\theta$ and $D_\theta^2\eta = -\eta$
while $\partial^2_\theta z = -z$.
}
:
\be
{\partial^2 z\over\partial \tau^2}=-{R\over 12}z
\ee

This equation, in turn,
can be converted into the dispersion relation
\be\label{osc}
{\cal E} = \frac{p^2}{2} + \frac{R}{24}z^2\,,
\ee
where $p$ is the momentum conjugate to $z$.
It is amusing to notice, that this equation describes a non-relativistic inverted harmonic oscillator
and the tunneling probability is again given by the integral between the turning points
and is proportional to (after putting ${\cal E}=-2m$)
\be
\label{prob}
\exp\left(-2\int_0^l |p(z)|d z\right) =\theta(-R)
\exp\left(-2\int_0^l\sqrt{4m + {R\over 12}z^2}\ dz\right)
= \theta (-R)\exp\left(-\frac{2\pi m}{\sqrt{-R/12}}\right)\,.
\ee
Again, the applicability of the quasiclassical approximation used here, requires that the exponent
in this formula is large, i.e. $m^2\gg -R/12\equiv H^2$. 

This formula implies that particles are produced
in de Sitter space when the Ricci scalar is negative, $R=-12H^2<0$ ($H$ is the Hubble constant)
with probability per unit volume and time interval given by $\exp\left(-{2\pi m / H}\right)$ \cite{Prob},
while they  are not produced in AdS space, where the Ricci scalar has the opposite sign and
the relevant geodesics decelerate rather than accelerate
\footnote{The emergence of the oscillator potential (\ref{osc})
in the problem related to de Sitter space should
 not come as a surprise: in the static coordinate system the de Sitter metric
has the form $$ds^2=(1-H^2 r^2) dt^2 - \frac{1}{1-H^2 r^2} dr^2 - r^2 d\Omega_2^2,$$ which in the Newtonian approximation corresponds to the inverted oscillator potential $V=-H^2 r^2/2$.
}
.
In the case of massless gravitons in de Sitter, the role of $m^2$ is played by $k^2/a^2(t)$, for a given mode
with comoving momentum $k$. The production of modes with $k^2/a^2(t) \lesssim H^2$ is expected not to be
exponentially suppressed \cite{LPP}.

\vspace{0.5cm}

{\bf 3.} Similarly, one can deal with more complicated non-maximally symmetric spacetimes.
In order to determine if particles are
produced in these cases, it is not sufficient to look at the sign of the
curvature scalar; instead, one has to proceed with a more refined analysis of the geodesic deviation equation.
For instance, consider the Friedmann-Robertson-Walker (FRW) metric
\be
ds^2=\Omega^2(x^0)\left((dx^0)^2-(d\vec x)^2\right)=dt^2-a^2(t)(d\vec x)^2
\ee
and assume for simplicity that the 3-dimensional space is flat. Then,
the  curvature scalar is equal to
\be
R=-6\frac{\Omega''}{\Omega^3}\,,
\ee
where the prime denotes differentiation with respect to $x^0$, and
the geodesic deviation equation reads
\be
\label{genc}
D_\tau^2\eta^{\,i}=\left(\frac{\Omega''}{\Omega}-2\frac{{\Omega'}^2}{\Omega^2}\right)
\left((u^0)^2\eta^{\,i} -(u^0\eta^0) u^{\,i}\right)+ \frac{{\Omega'}^2}{\Omega^4}
\left(\eta^{\,i}-(\eta u)u^{\,i}\right) \,.
\ee
In the case of de Sitter, $\Omega(x^0)=1/(Hx^0)=\exp (Ht)$. Thus,
the first term in this equation vanishes and one returns to
formula (\ref{ddeta1}). In the generic FRW case, on the other hand,
in the comoving coordinates of the central geodesic i.e. for $u^\mu=(1,0,0,0)$, equation (\ref{genc})
takes the form
\be
D^2\eta^{\,i}=\left(\frac{\Omega''}{\Omega^3}-
\frac{{\Omega'}^2}{\Omega^4}\right)\eta^{\,i}=\frac{\ddot a}{a}\eta^{\,i}\,,
\ee
where the dots denote differentiation with respect to $t$. Again, this equation
can be rewritten as an equation for the distance $z$ between two neighboring geodesics, with the covariant
derivatives replaced by the ordinary ones with respect to $t$, namely
\be\label{mde}
{\partial^2 z\over\partial t^2}=\frac{\ddot a}{a} z\,.
\ee
This equation also describes an oscillator but with
time-dependent frequency, and, depending on the explicit form of $a(t)$, there can be
relatively accelerating geodesics, in which case particle creation will take place.

The lesson is that the particle production is determined in general not simply by the sign of the curvature scalar,
but by the sign of $\ddot a$ (given that $a>0$). Pairs are produced only if
$\ddot a>0$. Note that a solution of (\ref{mde}) is just $z(t)=a(t)$.
For a power-law behavior of the scaling factor
$a(t)\sim t^s$, the condition $\ddot a>0$ becomes $s(s-1)>0$. This inequality is
satisfied for $s>1$ or $s<0$. The latter behaviour, however, describes a contracting universe in
which, of course, there is no chance of particle creation. On the contrary, in the case of
$s>1$ describing a superluminal accelerated expansion, pair production takes place.
In order to calculate the production rate in these FRW metrics, one has to deal with
an oscillator with varying frequency and, correspondingly, a time dependent barrier. 
This requires detailed analysis and the answer will depend on the explicit form of $a(t)$ \cite{ZP}.

\vspace{0.5cm}

{\bf 4.} A few simple thoughts related to the fate of de Sitter evolution as a result of particle production are in order.
Since emerging particles gravitate, they act against the ``anti-gravity"
of de Sitter space, which ``repels" particles. It is an old hypothesis \cite{Pol2},
that maybe the back reaction of created particles could drastically affect the fate of the de Sitter
evolution, perhaps even stopping its accelerated expansion.
For this to happen, however, the density of created particles should be big enough and maybe ``explosive".
This is not easy to achieve due to the dilution of the produced particles by the space expansion.

The particle density $n=N/V$ in the absence of self-interactions is governed by the well-known equation
\be
\label{ndot}
\dot n = \alpha - n\frac{\dot V}{V}\,,
\ee
where the first term on the r.h.s. describes the creation of particles, $\alpha\sim
\exp\left(-{2\pi m / H}\right)$, while the second term describes their dilution.
In the exponentially expanding  flat de Sitter one obtains
$V \sim \exp(3Ht)$, so that
$\dot V/V = 3H = {\rm constant}$ and (\ref{ndot})
implies that
\be
\label{keqsol}
n = \frac{\alpha}{3H} + \beta e^{-3Ht}\,,
\ee
which does not grow with time and the first term is exponentially small
so that the creation of particles {\it cannot}
stop the accelerated expansion.

This well-known result is due to the exponential expansion of de
Sitter space. One way one might expect to change the situation is by
introducing self-interaction among the produced particles (see
\cite{Pol,selfi} for further discussion). The self-interaction can
stimulate production of particles with a chain reaction,
represented in (\ref{ndot}) by an extra term of the form $\gamma n$
with $\gamma>0$ on the right hand side. Hence, if $\gamma >3H$, this
term dominates over dilution and  particle
production can eventually cancel the de Sitter exponential expansion, as a result of
the exponential increase of the matter density \footnote{If the spatial
section of the space-time is closed, i.e. $V\sim \cosh^3(Ht)$, one
obtains instead
\be
n={\alpha\over 8 \cosh^3 (Ht)}\left[
{e^{3Ht}\over 3H-\gamma}-{e^{-3Ht}\over 3H+\gamma} +{e^{Ht}\over
H-\gamma}-{e^{-Ht}\over H+\gamma}\right]+ \beta {e^{\gamma
t}\over\cosh^3(Ht)}
\ee
which does not change the conclusion. }:
\be
n = \frac{\alpha}{3H-\gamma} + \beta e^{(\gamma -3H)t}\,.
\ee
No quantum correlations, fusion of created particles, or the effect of
Bose statistics was taken into account above. The latter, in
particular may lead to higher powers of $n$ on the r.h.s. of the
density evolution equation and has been discussed in \cite{Pol}.

\vspace{0.5cm}

{\bf 5.} Any account of self-interaction which leads to fast
increase of particle density implies that the Einstein equations should be modified. In
particular, this means that the acceleration of the Universe expansion can change to deceleration. Indeed,
\be
\label{acc}
\frac{\ddot a}{a}\sim - (3p+\epsilon)
\ee
where $\epsilon$ is the energy density and $p$ is the pressure. Since in the de Sitter
space $\epsilon=H^2=-p$, the acceleration is positive. However, if enough particles are produced
with positive both $\epsilon$ and $p$ (the relation between
$\epsilon$ and $p$ depends on the model), one can change the sign of the r.h.s. of (\ref{acc}).

However, the curvature
\be
R \sim -(\epsilon - 3p)=-T^\mu_{\;\;\mu}
\ee
only {\it increases}
\footnote{
In \cite{WT}, the authors obtained a decrease of the cosmological constant
as a result of negative corrections to $\epsilon$ due to the interaction between
gravitons.}
in its absolute value
when any real particles are created,
with $\epsilon-3p \geq 0$.
Thus, while creation of particles can indeed act against and, perhaps, even stop
the accelerated expansion, looking at the curvature and assuming that it is the negative curvature which controls
particle production one could, at first sight, think that there is violation
of the Le Chatelier principle, namely, that negative curvature creates particles, and at the same time it
becomes more negative as a result of this. However, as we saw in Section 3, it is
the acceleration $\ddot a$ and not the curvature that controls the
particle production. Since, according to (\ref{acc}) the acceleration decreases as a result of particle creation,
there is no contradiction with the Le Chatelier principle.

\bigskip

{\bf 6.}
An elementary quantum mechanical (as opposed to quantum field theoretic)
discussion of particle
production in gravitational backgrounds was presented,
appealing as much as possible to physical intuition and carefully avoiding the 
formal and usually technically very involved argumentation. It relates the phenomenon of pair production
to the accelerated deviation of nearby geodesics in the given spacetime background. In simple
cases of maximally symmetric
backgrounds, it leads to the known exponential barrier penetration formula for the probability of such particle production.
The approach is general, valid in locally curved spacetimes of any dimensionality.
Our arguments are not expected to be new
to the experts, but surprisingly, they are not, to the best of our
knowledge, explicitly presented in the existing literature.

\bigskip

Work supported in part by the EU grant FP7-REGPOT-2008-1-CreteHEPCosmo-228644, as well as by the
grants of the Russian Federal Nuclear Energy Agency under contract H.4e.45.90.11.1059, by the
Ministry of Education and Science of the Russian Federation under contract 02.740.11.0081, by RFBR
grants 10-02-00509 (A. Mir.), 10-02-00499 (A. Mor.),
by joint grants 11-02-90453-Ukr, 09-02-93105-CNRSL, 09-02-91005-ANF,
10-02-92109-Yaf-a, 11-01-92612-Royal Society.
A. M. and A. M. would like to thank the Department of Physics and the ITCP of the University of Crete for their
hospitality and support during their visits.

\end{document}